\documentclass[%
 aip,
 jmp,
 amsmath,amssymb,
 reprint,onecolumn
]{revtex4-1}

\usepackage{graphicx}
\usepackage{bm}

\usepackage[utf8]{inputenc}
\usepackage[T1]{fontenc}
\usepackage{mathptmx}
\usepackage{etoolbox}

\makeatletter
\def\@email#1#2{%
 \endgroup
 \patchcmd{\titleblock@produce}
  {\frontmatter@RRAPformat}
  {\frontmatter@RRAPformat{\produce@RRAP{*#1\href{mailto:#2}{#2}}}\frontmatter@RRAPformat}
  {}{}
}%
\makeatother

\begin{document}

\preprint{AIP/123-QED}

\title{Interior spacetimes sourced by stationary differentially rotating irrotational cylindrical fluids. Perfect fluids}

\author{M.-N. C\'el\'erier}
\email{marie-noelle.celerier@obspm.fr}
\affiliation{Laboratoire Univers et Th\'eories, Observatoire de Paris, Universit\'e PSL, Universit\'e Paris Cit\'e, CNRS, F-92190 Meudon, France}

\date{\today}

\begin{abstract}

In a recent series of papers new exact analytical solutions of the Einstein equations representing interior spacetimes sourced by stationary rigidly rotating cylinders of different kinds of fluids  have been displayed, [Phys. Rev. D {\bf 104}, 064040 (2021); J. Math. Phys. {\bf 64}, 022501 (2023); J. Math. Phys. {\bf 64}, 032501 (2023); J. Math. Phys. {\bf 64}, 042501 (2023); and J. Math. Phys. {\bf 64}, 052502 (2023)]. This work is currently being extended to the cases of differentially rotating irrotational fluids. The results are presented in a new series of papers considering in turn the same three anisotropic pressure cases, as well as a perfect fluid source. Here, the perfect fluid case is considered, and different classes are identified as directly issuing from the field equations. Among them, an explicit analytical set of solutions is selected as displaying perfect fluid spacetimes. Its mathematical and physical properties are analyzed. Its matching to an exterior Lewis-Weyl vacuum and the conditions for avoiding an angular deficit are discussed.

\end{abstract}


\maketitle 

\onecolumngrid 
\section{Introduction} \label{intro}

The search for exact solutions of the Einstein field equations of General Relativity (GR) has been mainly focused on vacuum spacetimes since the finding of those describing the interior of a gravitational source is much more involved from both a mathematical and physical point of view \cite{G09,S09}. Therefore, some strongly simplifying assumptions are needed to be able to fully integrate interior solutions. 

In a series of articles devoted to the study of interior spacetimes sourced by fluids displaying different configurations of energy density and pressure \cite{CS20,C21,C23a,C23b,C23c,C23d,C23e}, these simplifying assumptions have been: cylindrical symmetry of the source implying two Killing vectors, stationary motion implying a third Killing vector, and rigid (shear-free) rotation allowing the use of corotating coordinates. A class of fully integrated solutions has been displayed for the case of a perfect fluid, i. e., with isotropic pressure \cite{C23a}. The anisotropic pressure configuration has been split into three different cases where the principal stresses are vanishing by pair and a number of classes and subclasses of exact solutions have been exhibited in this context \cite{C21,C23b,C23c,C23d}.

Here, and in the following papers of this new series, we make a step further and consider a differentially rotating fluid, therefore giving up the corotating coordinate simplification. This implies the occurence of a new quantity to be determined, the global rotation parameter $\Omega$, which increases the number of degrees of freedom of the problem. To overcome this impediment, the assumption of irrotational fluid is made. It allows in addition to simplify the equations to be solved.

In the present paper, the case of a perfect fluid is considered. Issuing directly from the field equations, two main classes of solutions are found, the second giving rise to five subclasses. Each class and subclass is thoroughly analyzed and a non-trivial subclass is sorted out. It corresponds to a fluid with an equation of state of the form $\rho= h P$, where $h$ is a constant. The equation of state is therefore polytropic, which is a feature usually welcome in astrophysics. However, applying junction conditions on the boundary of the cylinder of matter imposes that these spacetimes are either vacuum, or thread-like, or an infinitely wide cylinder, implying a somehow disturbing matching procedure. Moreover, the conditions for the avoidance of an angular deficit in the vicinity of the axis yield a negative energy density, therefore violating the weak energy condition. These results do not preclude the corresponding solutions to represent genuine GR spacetimes. They merely reduce their physical application domain.

The paper is organized as follows: after the introduction provided in present Sec. \ref{intro}, the statement of the problem from the most general issue to the irrotational perfect fluid configuration is depicted in Sec. \ref{sp}. The two main classes of solutions are identified in Sec. \ref{solver}. Class A is analyzed in Sec. \ref{classA}, and the five subclasses of class B are studied in Sec. \ref{classB}. A comparison of the features pertaining to the spacetimes sourced by differentially rotating perfect fluids studied here with those exhibited by the spacetimes generated by rigidly rotating perfect fluids considered formerly \cite{C23a} is presented in Sec. \ref{rr}. Section \ref{concl} is devoted to the conclusion.

\section{Statement of the problem} \label{sp}

\subsection{Description of the most general issue}

The most general gravitational source, whose different specializations will be considered in the series of works initiated here, is a stationary cylindrically symmetric fluid differentially rotating around its axis. It is bounded by a cylindrical hypersurface $\Sigma$. No equation of state being a priori imposed, its pressure is determined by its three principal stresses and follows from solving the field equations themselves. Its stress-energy tensor, displayed as (1) of C\'el\'erier and Santos \cite{CS20}, can been written as
\begin{equation}
T_{\alpha \beta} = (\rho + P_r) V_\alpha V_\beta + P_r g_{\alpha \beta}
+ (P_\phi - P_r) K_\alpha K_\beta + (P_z - P_r)S_\alpha S_\beta, \label{setens}
\end{equation}
where $\rho$ is the energy density of the fluid, $P_r$, $P_z$ and $P_\phi$ are the principal stresses. The four-velocity $V_\alpha$, and the four-vectors $K_\alpha$ and $S_\alpha$ satisfy
\begin{equation}
V^\alpha V_\alpha = -1, \quad K^\alpha K_\alpha = S^\alpha S_\alpha = 1, \quad V^\alpha K_\alpha = V^\alpha S_\alpha = K^\alpha S_\alpha =0. \label{fourvec}
\end{equation}

A spacelike hypersurface orthogonal Killing vector $\partial_z$ is assumed, such as to ease a subsequent proper junction to a vacuum exterior Lewis metric. Therefore, in geometric units $c=G=1$, the line element reads
\begin{equation}
\textrm{d}s^2=-f \textrm{d}t^2 + 2 k \textrm{d}t \textrm{d}\phi +\textrm{e}^\mu (\textrm{d}r^2 +\textrm{d}z^2) + l \textrm{d}\phi^2. \label{metric}
\end{equation}
The metric coefficients $f$, $k$, $\mu$, and $l$ are assumed to be real functions of the radial coordinate $r$ only, accounting therefore for stationarity. Owing to the cylindrical symmetry of the system, the coordinates must conform to the following ranges:
\begin{equation}
- \infty \leq t \leq +\infty, \quad 0 \leq r \leq +\infty, \quad -\infty \leq z \leq +\infty \quad 0 \leq \phi \leq 2 \pi, \label{ranges}
\end{equation}
the two limits $0$ and $2\pi$ of $\phi$ being topologically identified. The coordinates are denoted $x^0=t$, $x^1=r$, $x^2=z$, and $x^3=\phi$.

The four-velocity of the fluid, satisfying (\ref{fourvec}), is written as
\begin{equation}
V^\alpha = v \delta^\alpha_0 + \Omega  \delta^\alpha_3 , \label{4velocity}
\end{equation}
where the velocity $v$ and the global rotation $\Omega$ are functions of $r$ only. The timelike condition for $V^\alpha$ displayed in (\ref{fourvec}) becomes thus
\begin{equation}
fv^2 - 2kv\Omega - l\Omega^2 -1= 0. \label{timelike}
\end{equation}

The two spacelike four-vectors used to define the stress-energy tensor, and verifying conditions (\ref{fourvec}) can be chosen as
\begin{equation}
K^\alpha = -\frac{1}{D}\left[(kv+l\Omega)\delta^\alpha_0 + (fv - k\Omega)\delta^\alpha_3\right], \label{kalpha}
\end{equation}
\begin{equation}
S^\alpha = \textrm{e}^{-\mu/2}\delta^\alpha_2. \label{salpha}
\end{equation}

To allow the integration of the field equations, we have introduce the ansatz used in C\'el\'erier \cite{C21,C23a,C23b,C23c,C23d}, i. e., two key auxiliary functions of the radial coordinate $r$: $D(r)$ given by \cite{D06}
\begin{equation}
D^2 = fl + k^2, \label{D2}
\end{equation}
and the $h(r)$ function defined as \cite{C21}
\begin{equation}
h=\frac{P}{\rho}. \label{hdef}    
\end{equation}

\subsection{General field equations} \label{gfe}

Inserting (\ref{fourvec})--(\ref{hdef}) into (\ref{setens}), we obtain the components of the stress-energy tensor matching the nonvanishing components of the Einstein tensor. The five corresponding field equations can be written as \cite{CS20}
\begin{equation}
G_{00} = \frac{\textrm{e}^{-\mu}}{2} \left[-f\mu'' - 2f\frac{D''}{D} + f'' - f'\frac{D'}{D} + \frac{3f(f'l' + k'^2)}{2D^2}\right]
= \kappa \left[\rho f + (\rho + P_\phi)D^2\Omega^2\right], \label{gG00}
\end{equation}
\begin{equation}
G_{03} =  \frac{\textrm{e}^{-\mu}}{2} \left[k\mu'' + 2 k \frac{D''}{D} -k'' + k'\frac{D'}{D} - \frac{3k(f'l' + k'^2)}{2D^2}\right]
= - \kappa \left[\rho k + (\rho + P_\phi)D^2 v \Omega\right], \label{gG03}
\end{equation}
\begin{equation} 
G_{11} = \frac{\mu' D'}{2D} + \frac{f'l' + k'^2}{4D^2} = \kappa P_r \textrm{e}^\mu, \label{gG11}
\end{equation}
\begin{equation}
G_{22} = \frac{D''}{D} -\frac{\mu' D'}{2D} - \frac{f'l' + k'^2}{4D^2} = \kappa P_z \textrm{e}^\mu, \label{gG22}
\end{equation}
\begin{equation}
G_{33} = \frac{\textrm{e}^{-\mu}}{2} \left[l\mu'' + 2l\frac{D''}{D} - l'' + l'\frac{D'}{D} - \frac{3l(f'l' + k'^2)}{2D^2}\right]
= - \kappa \left[\rho l - (\rho + P_\phi)D^2 v^2\right], \label{gG33}
\end{equation}
where the primes stand for differentiation with respect to $r$.

\subsection{Junction conditions} \label{jc}

When applying a given solution to represent a standard astrophysical object, junction with an exterior vacuum spacetime is needed. The junction conditions have been worked out by Debbasch et al \cite{D06} for the case of a rigidly rotating cylinder of matter. The main condition, providing a powerful mean of discriminating physically well-behaved classes of solutions from others, i. e., $P_r =0$ on the boundary, has been widely used for the study of rigidly rotating fluids \cite{CS20,C21,C23a,C23b,C23c,C23d}. Its adaptation to differential rotation is now detailed here.

Since the motion is stationary, the vacuum exterior retained is once again the Lewis-Weyl spacetime \cite{L32} whose metric is recalled below as
\begin{equation}
\textrm{d}s^2=-F \textrm{d}t^2 + 2 K \textrm{d}t \textrm{d}\phi +\textrm{e}^M (\textrm{d}R^2 +\textrm{d}z^2) + L \textrm{d}\phi^2, \label{Wmetric}
\end{equation}
where
\begin{equation}
F= a R^{1 - n} - a \delta^2 R^{1 + n}, \label{W1}
\end{equation}
\begin{equation}
K = - (1 - ab\delta)\delta R^{1 + n} - ab  R^{1 - n}, \label{W2}
\end{equation}
\begin{equation}
\textrm{e}^M =  R^{(n^2 - 1)/2},  \label{W3}
\end{equation}
\begin{equation}
L = \frac{(1 - ab\delta)^2}{a} R^{1 + n} - ab^2 R^{1 - n}, \label{W4}
\end{equation}
with
\begin{equation}
\delta = \frac{c_{LW}}{an}, \label{W5}
\end{equation}
where $a, b,$ and $n$ are real constants, and $c_{LW}$ is set for the Lewis $c$ parameter, not to be confused with our own $c$ parameter.

In accordance with Darmois' junction conditions \cite{D27}, the coefficients of metric (\ref{metric}) and of metric (\ref{Wmetric}) and their derivatives must be continuous across the $\Sigma$ surface, which implies, from the continuity of the first fundamental form,
\begin{equation}
f  \stackrel{\Sigma}{= }a_1 F, \quad k  \stackrel{\Sigma}{=} a_2 K, \quad \textrm{e}^\mu  \stackrel{\Sigma}{=} a_3 \textrm{e}^M, \quad l  \stackrel{\Sigma}{=} a_4 L, \label{W6}
\end{equation}
where $\stackrel{\Sigma}{=}$ denotes that the values are taken at the boundary, and where the $a_1$, $a_2$, $a_3$ and $a_4$ constants can be transformed away by a rescaling of the coordinates, while from the continuity of the second fundamental form, we have
\begin{equation}
\frac{f'}{f}  \stackrel{\Sigma}{=} \frac{1}{R} + n \frac{\delta^2 R^n + R^{-n}}{\delta^2 R^{1 + n} - R^{1 - n}}, \label{W7}
\end{equation}
\begin{equation}
\frac{k'}{k}  \stackrel{\Sigma}{=} \frac{1}{R} + n \frac{(1 - ab\delta)\delta R^n - abR^{-n}}{(1 - ab\delta)\delta R^{1 + n} + abR^{1 - n}}, \label{W8}
\end{equation}
\begin{equation}
\mu' \stackrel{\Sigma}{=} \frac{n^2 - 1}{2R},  \label{W9}
\end{equation}
\begin{equation}
\frac{l'}{l}  \stackrel{\Sigma}{=} \frac{1}{R} + n \frac{(1 - ab\delta)^2 R^n + a^2b^2 R^{-n}}{(1 - ab\delta)^2 R^{1 + n} - a^2b^2R^{1 - n}}. \label{W10}
\end{equation}

The various conditions issuing from the above set of equations have been discussed at length by Debasch et al. \cite{D06} and C\'el\'erier and Santos \cite{CS20} in the rigid rotation case. We show below that the main constraint, useful for our purpose, is analogous in the differentially rotating case.

Indeed, in the most general case, the above equations inserted into (\ref{gG11}) imply $P_r \stackrel{\Sigma}{=} 0$, as expected. In the case of a perfect fluid, this condition must be imposed on the isotropic component of the pressure $P$ which plays the same role as $P_r$ in the ($G_{11}$) equation. This constraint will be implemented in the following to determine whether mathematically well-behaved classes of solution can be used in a standard astrophysical context.

The other junction conditions involve relations between the parameters defining each realization of the solutions for the interior and the exterior. Even though they are of no use to our analyses, they are briefly dealt with in Sec.\ref{mjc}.

\subsection{Irrotational field equations for a perfect fluid} \label{fe}

Now, the problem, when stated as in Sec. \ref{gfe}, and specialized to the perfect fluid case where $P_r=P_z=P_{\phi}=P$, is under-determined since we have a set of six equations, the five field differential equations plus the timelike condition, for eight unknowns, $f$, $\textrm{e}^{\mu}$, $k$, $l$, $P$, $\rho$ or alternatively $h$, $v$ and $\Omega$. We are therefore provided with two degrees of freedom which we propose to use in a manner allowing us to obtain a mathematically solvable and physically meaningful system.

Irrotational motion will be our choice to alleviate a first degree of indeterminacy. It implies that the local rotation tensor defined with respect to the four-velocity of the fluid and, in particular, its rotation scalar, vanishes. From the expression of this scalar given by (31) of C\'el\'erier and Santos \cite{CS20}, this vanishing yields
\begin{equation}
kv+l\Omega=0. \label{irr}    
\end{equation}

The second degree of freedom is presently kept aside to be used further on when we are led to discriminate between different classes of solutions to this problem.

Inserting (\ref{irr}) into (\ref{gG00})-(\ref{gG33}), we obtain the five corresponding field equations which can be written as
\begin{eqnarray}
-\mu'' - 2\frac{D''}{D} + \frac{f''}{f} - \frac{f'D'}{fD} + \frac{3(f' l' + k'^2)}{2D^2}
= \frac{2 \kappa}{fl} \left(D^2 \rho + k^2P\right) \textrm{e}^\mu, \label{G00}
\end{eqnarray}
\begin{equation}
\mu'' + 2 \frac{D''}{D} -\frac{k''}{k} + \frac{k'D'}{kD} - \frac{3(f' l' + k'^2)}{2D^2} = 2 \kappa P \textrm{e}^\mu, \label{G03}
\end{equation}
\begin{equation} 
\frac{\mu' D'}{D} + \frac{f' l' + k'^2}{2D^2} = 2 \kappa P \textrm{e}^{\mu}, \label{G11}
\end{equation}
\begin{equation}
\frac{2D''}{D} -\frac{\mu' D'}{D} - \frac{f'l' + k'^2}{2D^2} = 2 \kappa P \textrm{e}^{\mu} , \label{G22}
\end{equation}
\begin{equation}
\mu'' + 2\frac{D''}{D} - \frac{l''}{l} + \frac{l'D'}{lD} - \frac{3(f' l' + k'^2)}{2D^2}
= 2 \kappa P \textrm{e}^{\mu}. \label{G33}
\end{equation}

This choice of irrotational motion does indeed widely simplify these equations, since four among the five components of interest of the stress-energy tensor are now equal.

\subsection{Conservation of the stress-energy tensor} \label{bi}

The conservation of the stress-energy tensor is implemented by the Bianchi identity, whose general form is available as (57) in C\'el\'erier and Santos \cite{CS20}. Specialized to the present case and using the definition (\ref{hdef}) it becomes
\begin{equation}
\frac{h}{1+h}\frac{P'}{P} -\frac{l'}{2l} + \frac{D'}{D}= 0. \label{Bianchi}
\end{equation}

\subsection{Some useful equations} \label{rem}

The main key equations, established in previous works and still applying here with the $l$ metric function replacing $f$, are recalled below. That initially displayed as (14) in Debbasch et al. \cite{D06} and adapted here to the new problem, i. e., obtained from the subtraction of (\ref{G03}) from (\ref{G33}), can be written as
\begin{equation}
kl' - lk' = 2c D, \label{6}
\end{equation}
where $c$ is an integration constant and the factor 2 is added for further convenience. The $k$ solution of (\ref{6}) reads
\begin{equation}
k = l \left(c_k - 2c\int_{r_1}^r \frac{D(v)}{l(v)^2} \textrm{d}v \right). \label{7}
\end{equation}

With the help of (\ref{D2}), we obtain another useful relation, whose analogue has been established as (19) in C\'el\'erier \cite{C23b}, and which reads
\begin{equation}
\frac{f'l' +k'^2}{2D^2} = \frac{l'D'}{lD} - \frac{l'^2}{2l^2} + \frac{2c^2}{l^2}. \label{8}
\end{equation}

\subsection{More about the junction conditions} \label{mjc}

It is interesting to remark that the whole set of junction conditions, established in Sec. \ref{jc} when applied to an irrotational differentially rotating fluid, does not strictly reproduce that relevant to rigid rotation. Even though it will be of no use here, we display, as an example, the result obtained by substituting  (\ref{W6}), (\ref{W8}) and (\ref{W10}) into (\ref{6}), which reads
\begin{equation}
c = b(b c_{LW} - n).  \label{W11}
\end{equation}
This relation is different from the corresponding one in the rigid case, where both parameters $c$ and $c_{LW}$ were equal, allowing a straightforward interpretation of $c_{LW}$ \cite{C21,D06}, which is less easy here.

\section{Identifying two main classes of solutions} \label{solver}

A manipulation of the field equations yields a second degree equation for $D'/D$, which determines two different classes of solutions. It runs as follows.

Adding (\ref{G11}) and (\ref{G22}), we obtain
\begin{equation}
\frac{D''}{D} = 2 \kappa P \textrm{e}^{\mu}. \label{12}
\end{equation}
Now, (\ref{G11}) can be written as
\begin{equation}
\frac{f'l' +k'^2}{2D^2} = 2 \kappa P \textrm{e}^{\mu} - \frac{\mu'D'}{D}, \label{12a}
\end{equation}
which is inserted into (\ref{G33}), together with (\ref{12}), to give
\begin{equation}
\mu'' - \frac{l''}{l} - \frac{2D''}{D} + \left(3\mu' + \frac{l'}{l} \right) \frac{D'}{D} = 0. \label{12b}
\end{equation}

Equation (\ref{8}) where we insert (\ref{12a}), where (\ref{12}) has been previously substituted, becomes
\begin{equation}
\frac{D''}{D} = \left(\mu' + \frac{l'}{l} \right)\frac{D'}{D} - \frac{l'^2}{2 l^2}  + \frac{2 c^2}{l^2}, \label{13}
\end{equation}
which we insert into (\ref{12b}) to obtain
\begin{equation}
\frac{D'}{D} = - \frac{\left(\mu'' - \frac{l''}{l} + \frac{l'^2}{l^2} - \frac{4c^2}{l^2} \right)}{\mu' - \frac{l'}{l}}, \label{14}
\end{equation}
provided that $\mu' \neq l'/l$.

Now, we insert (\ref{12}) into (\ref{13}), that yields
\begin{equation}
2 \kappa P \textrm{e}^{\mu} = \left(\mu' + \frac{l'}{l} \right) \frac{D}{D}' - \frac{l'^2}{2l^2} + \frac{2c^2}{l^2}, \label{15}
\end{equation}
of which we take the logarithm, that we differentiate with respect to $r$. Inserting (\ref{Bianchi}) into the result, such as to get rid of $P'/P$, we obtain an equation involving $D''/D$ which we eliminate thanks to the use of (\ref{13}) to obtain
\begin{eqnarray}
&&\left[\frac{1+h}{h} \left(\frac{l'}{2l} - \frac{D'}{D}\right) + \mu' \right]\left[\left(\mu' + \frac{l'}{l} \right) \frac{D'}{D} - \frac{l'^2}{2l^2} + \frac{2c^2}{l^2} \right]
= \left(\mu'' + \frac{l''}{l} - \frac{l'^2}{l^2}\right)\frac{D'}{D} + \left(\mu' + \frac{l'}{l} \right) \left[\left(\mu' + \frac{l'}{l} \right) \frac{D'}{D} - \frac{l'^2}{2l^2}  \right. \nonumber \\
&+&  \left. \frac{2c^2}{l^2} - \frac{D'^2}{D^2}\right]- \frac{l'l''}{l^2} + \frac{l'^3}{l^3} - \frac{4c^2 l'}{l^3}, \label{16}
\end{eqnarray}
where we insert (\ref{14}) such as to be left with a relation between $\mu$, $l$ and their first and second derivatives with respect to $r$, which we multiply by $(\mu' - l'/l)^2$, hence recovering the possibility of considering $\mu' - l'/l = 0$ since this expression has thus disappeared from the denominator.

After some arrangements, we obtain
\begin{eqnarray}
&-& \left\{\frac{1+h}{h} \left[\left(\mu' -\frac{l'}{l}\right) \frac{l'}{2l} + \mu'' - \frac{l''}{l} + \frac{l'^2}{l^2} - \frac{4c^2}{l^2}\right] + \mu'\left(\mu' - \frac{l'}{l} \right)\right\} \left[\left(\mu' + 
\frac{l'}{l} \right)\left(\mu'' - \frac{l''}{l} + \frac{l'^2}{l^2} - \frac{4c^2}{l^2}\right) + \left(\mu' - \frac{l'}{l} \right)\left(\frac{l'^2}{2l^2} - \frac{2c^2}{l^2}\right) \right] \nonumber \\
&+& \left(\mu' - \frac{l'}{l} \right)\left(\mu'' + \frac{l''}{l} - \frac{l'^2}{l^2}\right)\left(\mu'' - \frac{l''}{l} + \frac{l'^2}{l^2} - \frac{4c^2}{l^2}\right) + \left(\mu' + 
\frac{l'}{l} \right)\left[\left(\mu' - \frac{l'}{l} \right) \left(\mu' + \frac{l'}{l} \right)\left(\mu'' - \frac{l''}{l} + \frac{l'^2}{l^2} - \frac{4c^2}{l^2}\right) \right.\nonumber \\
&+& \left.\left(\mu' - \frac{l'}{l} \right)^2 \left(\frac{l'^2}{2l^2} - \frac{2c^2}{l^2}\right)
+ \left(\mu'' - \frac{l''}{l} + \frac{l'^2}{l^2} - \frac{4c^2}{l^2}\right)^2\right] + \left(\mu' - \frac{l'}{l} \right)^2\left(\frac{l'l''}{l^2} - \frac{l'^3}{l^3} + \frac{4c^2l'}{l^3}\right) = 0. \label{17}
\end{eqnarray}

Now, we define a $g(r)$ function by
\begin{equation}
\mu' = g \frac{l'}{l}. \label{18}
\end{equation}
By inserting (\ref{18}) and its derivative into (\ref{12b}), then substituting the result into (\ref{13}), we obtain
\begin{equation}
g' \frac{l'}{l} + (g-1)\frac{l''}{l} - (g-1)\frac{l'^2}{l^2} + (g-1)\frac{l'D'}{lD}= \frac{4c^2}{l^2}. \label{19}
\end{equation}
Then, we insert (\ref{18}) and (\ref{19}) into (\ref{16}), which gives a second degree equation in $D'/D$ that reads
\begin{eqnarray}
&-&2(1-h+3g+hg)\frac{D'^2}{D^2} + \left[(1-h+5g+7hg) \frac{l'}{l} - 2(1+3h)g' + 2(1-h-g-3hg) \frac{l''}{l'}\right] \frac{D'}{D} + (-1+h+g+3hg) \frac{l''}{l} \nonumber \\
&-& g(1+3h) \frac{l'^2}{l^2} + (1+3h)\frac{g'l'}{l} = 0, \label{20}
\end{eqnarray}
whose solutions are
\begin{eqnarray}
\frac{D'}{D} = \frac{(1-h+5g+7 h g)\frac{l'}{l} - 2(1+3h)g' + 2(1-h-g-3 h g) \frac{l''}{l'}}{4(1-h+3g+ h g)} \nonumber \\
\pm \frac{(1-h+g-5 h g)\frac{l'}{l}+ 2 (1+3h)g' -2(1-h-g-3 h g) \frac{l''}{l'}}{4(1-h+3g+ h g)}. \label{21}
\end{eqnarray}

We have therefore identified two different classes of solutions: Class A defined by the $\pm$ sign in (\ref{21}) being a plus sign, and Class B, by it being a minus sign. The individualization of each class that we perform below amounts to using the last degree of freedom left for a full characterization of the problem, which is thus exactly determined.

\section{Class A} \label{classA}

This class is defined by choosing the $\pm$ sign in (\ref{21}) to be $+$, which gives
\begin{equation}
\frac{D'}{D} = \frac{l'}{2l},\label{22}
\end{equation}
which can be integrated by
\begin{equation}
D = c_D\sqrt{l},\label{23}
\end{equation}
where $c_D$ is an integration constant.

By substituting (\ref{22}) into (\ref{Bianchi}), we obtain
\begin{equation}
P' = 0,\label{23.a}
\end{equation}
which yields $P= const.$ Now, since the junction condition implies $P_{\Sigma}=0$ on the cylinder boundary, a constant pressure must vanish everywhere to satisfy this constraint. Class A is therefore a dust specialization of the perfect fluid solutions.

Now, by inserting $P=0$ into (\ref{12}), we obtain
\begin{equation}
D'' = 0,\label{24}
\end{equation}
which can be integrated as
\begin{equation}
D = c_D r+c_2,\label{25}
\end{equation}
where $c_D$ and $c_2$ are integration constants.

Then, we substitute (\ref{25}) into (\ref{22}) which becomes
\begin{equation}
\frac{l'}{l} = \frac{2  c_D}{c_D r+c_2},\label{25.a}
\end{equation}
which can be integrated by
\begin{equation}
l = c_3^2 (c_D r+c_2)^2,\label{26}
\end{equation}
where $c_3$ is another integration constant.

Now, by applying the axisymmetry condition \cite{C21,C23a,C23b,C23c,C23d,C23e}, $l\stackrel{0}{=} 0$ - where $\stackrel{0}{=}$ denotes that the value is taken at the axis - to (\ref{26}), we obtain $c_2=0$ and thus
\begin{equation}
D = c_D r,\label{27}
\end{equation}
and
\begin{equation}
l = c_l^2 r^2,\label{28}
\end{equation}
where we have renamed $c_3 c_D=c_l$.

Then, by inserting (\ref{27}) and (\ref{28}) into (\ref{7}), and integrating we obtain
\begin{equation}
k = c_l^2 c_k r^2 + \frac{c c_D}{c_l^2}.\label{29}
\end{equation}

The $f$ metric function results from the definition of $D$, given by (\ref{D2}), as
\begin{equation}
f = \frac{c_D^2}{c_l^2} - \left(c_l c_k r + \frac{c c_D}{c_l^3 r}\right)^2.\label{30}
\end{equation}

Now, we substitute $P=0$, and (\ref{27})-(\ref{30}) into the field equation (\ref{G11}) and obtain
\begin{equation}
\mu' = - \frac{2c^2}{c_l^4 r^3},\label{31}
\end{equation}
which can be integrated by
\begin{equation}
\mu = \frac{c^2}{c_l^4 r^2},\label{32}
\end{equation}
which can be written as
\begin{equation}
\textrm{e}^{\mu} = \exp \left(\frac{c^2}{c_l^4 r^2}\right).\label{33}
\end{equation}

Then, we substitute $P=0$, (\ref{27})-(\ref{30}) and (\ref{33}) into (\ref{G00}) and obtain
\begin{equation}
\rho = - \frac{4c^2}{\kappa c_l^4 r^4}\exp \left(-\frac{c^2}{c_l^4 r^2}\right).\label{34}
\end{equation}
This expression for the energy density being negative definite, this solution does not verify the weak energy condition.

Anyhow, in case a dust system with negative energy density should be needed for some applications, we display below the expressions for the global rotation parameter and the fluid velocity whose expressions issued from (\ref{timelike}) and (\ref{irr}) are given here by
\begin{equation}
\Omega^2 = \frac{k^2}{l D^2}, \label{33a}
\end{equation}
\begin{equation}
v = - \frac{l \Omega}{k},\label{33b}
\end{equation}
which yield
\begin{equation}
\Omega = \frac{c_l c_k}{c_D} + \frac{c}{c_l^3 r^2}, \label{34.a}
\end{equation}
\begin{equation}
v = - \frac{c_l}{c_D}, \label{35}
\end{equation}
where the choice of a plus sign for $\Omega$, implying a minus sign for $v$, owing to (\ref{33b}), is arbitrary. The important feature is that these signs are opposite. 

Note that the global rotation parameter $\Omega$ never vanishes. This shows that the dust cylinder is actually differentially rotating.

However, this parameter diverges for $r=0$. This might denote a singular axis since the metric functions $f$ and $\textrm{e}^{\mu}$ diverge also at this location. Added to the failure to comply with the weak energy condition, this may worsen the poor behaviour displayed by class A and impair its possible astrophysical applications.

\section{Class B} \label{classB}

This class is defined by choosing the $\pm$ sign in (\ref{21}) to be $-$, which yields
\begin{equation}
\frac{D'}{D} = \frac{-(1+3h)\mu'' + (1-h)\frac{l''}{l}}{(1-h)\frac{l'}{l} + (3+h)\mu'}, \label{24a}
\end{equation}
which we equalize to (\ref{14}) and obtain
\begin{equation}
\left(\frac{2c^2}{l^2} - \frac{l'^2}{2l^2}\right)\left[(3+h)\mu' + (1-h)\frac{l'}{l} \right] =  (1-h)\mu'\mu'' +(1+h)\frac{\mu''l'}{l}-(1+h)\frac{\mu'l''}{l} - (1-h)\frac{l'l''}{l^2}, \label{25a}
\end{equation}
which we insert into (\ref{17}) that becomes
\begin{equation}
-4(1-h)(1+h)\mu'^2 \mu''\frac{l'l''}{l^2} = 0. \label{26a}
\end{equation}
Since any of the terms in the product can be set to zero, six subclasses of Class B emerge, each corresponding to the vanishing of one of the factors in (\ref{26a}).

\subsection{Subclass B.1}

This subclass is defined by the vanishing of $l'$, which implies $l=c_l= const.$

Now, the axisymmetry condition $l\stackrel{0}{=} 0$, yields $l = c_l=0$. Subclass B.1 is therefore ruled out.

\subsection{Subclass B.2} \label{B.2}

This subclass is obtained for $l''=0$ that can be integrated by
\begin{equation}
l = c_l r + c_1, \label{2.1}
\end{equation}
$c_l$ and $c_1$ being integration constants. The axisymmetry condition, recalled above, implies $c_1=0$. Thus (\ref{2.1}) becomes
\begin{equation}
l = c_l r. \label{2.2}
\end{equation}

Now, we substitute (\ref{2.1}), and derivatives, into the two independent Eqs. (\ref{13}), and (\ref{14}), such as to obtain
\begin{equation}
\frac{D''}{D} = \left(\mu' + \frac{1}{r} \right)\frac{D'}{D} - \frac{1}{2 r^2}  + \frac{2 c^2}{c_l^2 r^2}, \label{2.3}
\end{equation}
\begin{equation}
\frac{D'}{D} = - \frac{\left(\mu'' + \frac{1}{r^2} - \frac{4c^2}{c_l^2 r^2} \right)}{\mu' - \frac{1}{r}}. \label{2.4}
\end{equation}
Then, (\ref{2.1}) is inserted into (\ref{24a}), which becomes
\begin{equation}
\frac{D'}{D} = \frac{-(1+3h)\mu''}{(1-h)\frac{1}{r} + (3+h)\mu'}. \label{2.5}
\end{equation}

Now, we equalize (\ref{2.4}) and (\ref{2.5}) and obtain
\begin{equation}
2(1+h) \frac{\mu''}{r} + 2(1-h) \mu' \mu''+ \left(1-\frac{4c^2}{c_l^2}\right)\frac{(1-h)}{r^3} + \left(1-\frac{4c^2}{c_l^2}\right) (3+h) \frac{\mu'}{r^2} = 0. \label{2.6}
\end{equation}
It is easy to see that this first order differential equation in $\mu'$ possesses an analytical solution only provided that $h=const.$ We therefore confine ourselves, for the present subclass B.2 of solutions, to this particular kind of equation of state which is, however, widely used for different physical purposes. The solution to (\ref{2.6}) reads therefore
\begin{equation}
\mu' = \frac{a}{r}, \label{2.7}
\end{equation}
where $a$ is an integration constant. This equation can be integrated by
\begin{equation}
\textrm{e}^{\mu}= c_{\mu}r^a, \label{2.8}
\end{equation}
where $c_{\mu}$ can be set to unity by a proper rescaling of the $r$ and $z$ coordinates. Hence, we obtain
\begin{equation}
\textrm{e}^{\mu}= r^a. \label{2.9}
\end{equation}
By inserting (\ref{2.7}), and derivative, into (\ref{2.6}), we obtain
\begin{equation}
h = \frac{a-2a^2+1-(1+3a)\frac{4c^2}{c_l^2}}{a-2a^2+1-(1-a)\frac{4c^2}{c_l^2}}. \label{2.10}
\end{equation}

Now, we substitute (\ref{2.7}), and derivative, into (\ref{2.4}), which yields
\begin{equation}
\frac{D'}{D} = \left[1 + \frac{4c^2}{(a-1)c_l^2} \right] \frac{1}{r}, \label{2.11}
\end{equation}
which can be integrated by
\begin{equation}
D = c_D r^{1 + \frac{4c^2}{(a-1)c_l^2}}, \label{2.12}
\end{equation}
where $c_D$ is another integration constant.

Then, by inserting (\ref{2.2}) and (\ref{2.12}) into (\ref{7}), and integrating, the $k$ metric function emerges as
\begin{equation}
k = c_l r \left[c_k + \frac{c_D(1-a)}{2c} r^{ \frac{4c^2}{(a-1)c_l^2}}\right]. \label{2.14}
\end{equation}
The last metric function $f$ is readily obtained through (\ref{D2}) and reads
\begin{equation}
f = \frac{c_D^2}{c_l} r^{1 + \frac{8c^2}{(a-1)c_l^2}} - c_l r \left[c_k + \frac{c_D(1-a)}{2c} r^{ \frac{4c^2}{(a-1)c_l^2}}\right]^2. \label{2.15}
\end{equation}

Now, we substitute $l'/l=1/r$, from (\ref{2.2}), and (\ref{2.11}) into the Bianchi identity (\ref{Bianchi}) and obtain
\begin{equation}
\frac{P'}{P} = -\left(\frac{1+h}{h}\right) \left[\frac{1}{2} + \frac{4c^2}{(a-1)c_l^2} \right] \frac{1}{r}, \label{2.16}
\end{equation}
which can be integrated by
\begin{equation}
P = \frac{c_P}{r^{\left(\frac{1+h}{h}\right)\left[\frac{1}{2} + \frac{4c^2}{(a-1)c_l^2}\right]}}, \label{2.17}
\end{equation}
where $c_P$ is another integration constant. Note that, provided that $h \neq -1$ and $\frac{1}{2} + \frac{4c^2}{(a-1)c_l^2} \neq 0$, the pressure $P$ is not a constant. Therefore, for the widest range of parameters, this class of solutions corresponds to a perfect fluid with non vanishing pressure.

The energy density $\rho$ follows straightforwardly from $\rho=P/h$.

Finally, the rotation parameter $\Omega$ and the velocity of the fluid $v$ are obtained by inserting (\ref{2.2}), (\ref{2.12}) and (\ref{2.14}) into (\ref{33a}) and (\ref{33b}) which gives
\begin{equation}
\Omega^2 = \frac{c_l \left[c_k + \frac{c_D(1-a)}{2c} r^{ \frac{4c^2}{(a-1)c_l^2}}\right]^2}{c_D^2 r^{1 + \frac{8c^2}{(a-1)c_l^2}}}, \label{2.18}
\end{equation}
\begin{equation}
v = - \frac{\sqrt{c_l}}{c_D}\frac{1}{r^{\frac{1}{2} + \frac{4c^2}{(a-1)c_l^2}}}. \label{2.19}
\end{equation}
Note that, since $\Omega$ is not constant, the rotation is actually nonrigid.

\subsubsection{Constraints on the parameters from the field equations}

At this stage of the calculations, the solution is given in terms of six integration constants: $a$, $c$, $c_l$, $c_D$, $c_k$, and $c_P$, plus an equation of state parameter, $h$. However, they are not fully independent as shown by the relations displayed below.

We insert first (\ref{2.9}), (\ref{2.12}), and (\ref{2.17}) into (\ref{12}), itself issued directly from the field Eqs. (\ref{G11}) and (\ref{G22}), and obtain two constraint equations:
\begin{equation}
c_P = \frac{2 c^2}{\kappa (a-1)c_l^2}\left[1+ \frac{4c^2}{(a-1)c_l^2}\right], \label{2.20}
\end{equation}
and
\begin{equation}
h = \frac{\frac{1}{2} + \frac{4 c^2}{(a-1)c_l^2}}{a +\frac{3}{2} - \frac{4 c^2}{(a-1)c_l^2}}. \label{2.20a}
\end{equation}

Now, we substitute (\ref{2.7}), (\ref{2.11}), and derivative into (\ref{2.3}) so as to obtain a constraint equation relating parameters $a$, and $c^2/c_l^2$, which reads
\begin{equation}
(1+2a)(1-a)^2 + 4 (1-a)(1-3a)\frac{c^2}{c_l^2} - 32\frac{c^4}{c_l^4} =0, \label{2.21}
\end{equation}
which is a second degree equation in $c^2/c_l^2$ whose solutions are
\begin{equation}
\frac{c^2}{c_l^2} = \frac{(1-a)}{16} \left(1-3a + \epsilon \sqrt{9+10a+9a^2}\right), \label{2.22}
\end{equation}
with $\epsilon = \pm1$.

Then, by inserting (\ref{2.22}) into (\ref{2.20}) and (\ref{2.20a}), we obtain
\begin{equation}
c_P = \frac{1}{16 \kappa} \left[3 + 8a + 9a^2 - \epsilon(1+3a) \sqrt{9+10a+9a^2}\right], \label{2.22a}
\end{equation}
\begin{equation}
h = \frac{1+3a - \epsilon \sqrt{9+10a+9a^2}}{7 + a + \epsilon \sqrt{9+10a+9a^2}}. \label{2.22b}
\end{equation}

Owing to the three constraint Eqs. (\ref{2.22})-(\ref{2.22b}), the solutions of the field equations pertaining to class B.2 depend, therefore, on only four independent parameters which can be chosen at will among the seven initial ones. In practice, the equation of state parameter $h$ should be chosen, and therefore (\ref{2.22b}) will have to be solved for $a(h)$ which should then be inserted into (\ref{2.22}) and (\ref{2.22a}), so that the four free parameters should be $h$, $c_k$, $c$, or alternatively $c_l$, and $c_D$.

\subsubsection{Metric signature}

We have previously imposed to the solutions studied in the framework of rigid rotation \cite{C21,C23a,C23b,C23c,C23d} that the four metric functions should be all positive definite or all negative definite, so as to obtain a proper Lorentzian signature for the metric. However, this prescription is sufficient but not necessary, since the metric is not diagonal. Indeed, metric (\ref{metric}) can be converted to a diagonal-like form through a mere algebraic arrangement, i. e., without any change of coordinate system \cite{D96}. It thus becomes
\begin{equation}
\textrm{d}s^2=-f (kf^{-1}\textrm{d}\phi - \textrm{d}t)^2 + (l+k^2f^{-1}) \textrm{d}\phi^2 +\textrm{e}^\mu (\textrm{d}r^2 +\textrm{d}z^2), \label{metric2}
\end{equation}
which implies a correct signature provided that $f$ and $l+k^2f^{-1}$ are both $>0$, since $\textrm{e}^\mu$ is indeed positive by virtue of $r$ being positive.

To begin with, we assume that the condition $f>0$ is fulfilled, and we consider the second inequality
\begin{equation}
l+\frac{k^2}{f} >0. \label{2.23}    
\end{equation}
The left-hand side, where we insert (\ref{2.2}), (\ref{2.14}), and (\ref{2.15}), can be written as
\begin{equation}
l+\frac{k^2}{f} = c_l r \left\{1 + \frac{1}{\frac{\frac{c_D^2}{c_l^2} r^{\frac{8c^2}{(a-1)c_l^2}}}{\left[c_k + \frac{c_D(1-a)}{2c} r^{ \frac{4c^2}{(a-1)c_l^2}}\right]^2} -1}\right\}. \label{2.24}    
\end{equation}

Two possibilities can occur.

a) Either $c_l>0$, in which case $l>0$ and the inequality (\ref{2.23}), is fulfilled, since $f$ is assumed $>0$.

b) Or $c_l<0$, and a straightforward analysis shows that (\ref{2.23}) is verified provided that
\begin{equation}
\frac{c_D^2}{c_l^2} r^{\frac{8c^2}{(a-1)c_l^2}}{\left[c_k + \frac{c_D(1-a)}{2c} r^{ \frac{4c^2}{(a-1)c_l^2}}\right]^2} >0, \label{2.25}    
\end{equation}
which is true since every term in this expression is $>0$. Hence, (\ref{2.23}) is satisfied whatever the sign of $l$. For this, we have assumed the metric function $f$ to be positive. It is the case provided that
\begin{equation}
\frac{c_D^2}{c_l} r^{1 + \frac{8c^2}{(a-1)c_l^2}} - c_l r \left[c_k + \frac{c_D(1-a)}{2c} r^{ \frac{4c^2}{(a-1)c_l^2}}\right]^2 >0,  \qquad \forall r. \label{2.26} 
\end{equation}
The fulfilment of this inequality depends on the values of five parameters, among which four are independent. The display of a thorough study of the different cases involved would be long and fastidious. Therefore, we leave it to be completed in the future for each possible application of these results, and we assume in the following that the considered spacetimes are those whose set of parameters allows a proper signature for the metric.

\subsubsection{Junction conditions}

As recalled in Sec. \ref{jc}, it is usual, for astrophysical applications, that interior solutions should be matched to an exterior vacuum. In the present case, we have seen that the proper vacuum to be considered is the Weyl class of the Lewis solutions generated by a stationary rotating cylinder of matter \cite{L32}. A proper matching to such a vacuum can occur, provided that $P_{\Sigma}=0$, where the index $\Sigma$ denotes values taken at the boundary.

Therefore, (\ref{2.17}) imposes
\begin{equation}
\frac{c_P}{r_{\Sigma}^{\left(\frac{1+h}{h}\right)\left[\frac{1}{2} + \frac{4c^2}{(a-1)c_l^2}\right]}} =0, \label{2.27}
\end{equation}
which implies that

a) $P_{\Sigma}=0$ for $r_{\Sigma} \rightarrow \infty$ provided that
\begin{equation}
\left(\frac{1+h}{h}\right)\left[\frac{1}{2} + \frac{4c^2}{(a-1)c_l^2}\right] > 0. \label{2.28}
\end{equation}

b) $P_{\Sigma}=0$ for $r_{\Sigma} =0$ provided that
\begin{equation}
\left(\frac{1+h}{h}\right)\left[\frac{1}{2} + \frac{4c^2}{(a-1)c_l^2}\right] < 0. \label{2.29}
\end{equation}

c) $P_{\Sigma}=c_P=0$ for
\begin{equation}
\left(\frac{1+h}{h}\right)\left[\frac{1}{2} + \frac{4c^2}{(a-1)c_l^2}\right] = 0, \label{2.30}
\end{equation}
which occurs either for $h=-1$ or for the vanishing of the expression in square brackets We will see in the following that, for $h=-1$, which is the case for class B.6, the solution is indeed $P= - \rho =0$ and is therefore a vacuum. Now this is also the case when
\begin{equation}
\left[\frac{1}{2} + \frac{4c^2}{(a-1)c_l^2}\right] = 0. \label{2.31}
\end{equation}
Using (\ref{2.21}), it is easy to show that this is the case when $a=-2$ and that the sign of the expression in (\ref{2.31}) depends not only of the respective values of $h$ and $a$, but also of that of $\epsilon$ in (\ref{2.22}). If the pressure vanishes for $r_{\Sigma}=0$, the cylinder reduces to an infinite thread of zero width, and its interior has no physical meaning. This case is therefore ruled out. When it vanishes for $r_{\Sigma} \rightarrow \infty$ the solution represents an infinitely wide cylinder, {\it a priori} devoid of pure astrophysical interest. However, we will analyze it further as an exact solution for the interior of a differentially rotating perfect fluid.

These results are summarized in two tables: table \ref{table1} displays them for $\epsilon = +1$, while table \ref{table2} displays them for $\epsilon = -1$.

\begin{table}  
\begin{tabular}{|c|c|c|c|c|}
\hline \multicolumn{5} {|c|} {Table I. $\epsilon = +1$} \\ \hline     & $ \hspace{0.5cm} h<-1 \hspace{0.5cm}$  & $ \hspace{0.5cm} h=-1 \hspace{0.5cm}$ & $ \hspace{0.5cm} -1<h<0 \hspace{0.5cm}$ & $ \hspace{0.5cm}0<h \hspace{0.5cm}$ \\ \hline \hspace{0.3cm} $a<-2$ \hspace{0.3cm} & Thread-like &  Vacuum  &  $\infty$ly wide & Thread-like \\ \hline $a=-2$  & Vacuum & Vacuum & Vacuum & Vacuum\\  \hline  $a>-2$ & Thread-like & Vacuum & $\infty$ly wide & Thread-like\\  \hline
\end{tabular}
\caption{\label{Table I} Nature of the spacetimes according to the crossed values of the parameters $h$ and $a$, when $\epsilon=+1$.}
\label{table1}
\end{table}

\begin{table}  
\begin{tabular}{|c|c|c|c|c|}
\hline \multicolumn{5} {|c|} {Table II. $\epsilon = -1$} \\ \hline     & $ \hspace{0.5cm} h<-1 \hspace{0.5cm}$  & $ \hspace{0.5cm} h=-1 \hspace{0.5cm}$ & $ \hspace{0.5cm} -1<h<0 \hspace{0.5cm}$ & $ \hspace{0.5cm}0<h \hspace{0.5cm}$ \\ \hline \hspace{0.3cm} $a<-2$ \hspace{0.3cm} & Thread-like &  Vacuum &  $\infty$ly wide & Thread-like \\ \hline $a=-2$  & Vacuum & Vacuum & Vacuum & Vacuum\\  \hline  $a>-2$ & $\infty$ly wide & Vacuum & Thread-like & $\infty$ly wide\\ \hline
\end{tabular}
\caption{\label{Table II} Nature of the spacetimes according to the crossed values of the parameters $h$ and $a$, when $\epsilon=-1.$}
\label{table2}
\end{table}

\subsubsection{Non-angular deficit condition}

The status of the so-called ``regularity'' condition is somehow complicated. We have shown, through particular examples, in the series of papers devoted to the study of rigidly rotating cylindrically symmetric spacetimes \cite{C21,C23a,C23b,C23c,C23d}, that this ``regularity'' condition, supposedly able to explain out any hidden singularity on the axis, is neither necessary nor sufficient for this purpose. It is rather a non-angular deficit condition. Actually, the method proposed by Mars and Senovilla \cite{M93} consists in imposing that the ratio of the circumference over the radius of an infinitesimally small circle around the rotation axis does not depart from the value $2 \pi$. This circle is defined as the orbit of the spacelike Killing vector $\vec{\xi}$ generating the axial symmetry isometry. This Killing vector must therefore satisfy 
\begin{equation}
\frac{\partial_{\alpha}\left(\xi^{\mu} \xi_{\mu}\right) \partial ^{\alpha}\left(\xi^{\nu} \xi_{\nu}\right)}{4 \xi^{\lambda}\xi_{\lambda}} \stackrel{0}{=} 1. \label{2.32}
\end{equation} 
Therefore, this so-called ``regularity'' condition \cite{S09} merely insures that no angular deficit occurs in the vicinity of the axis, e. g., no conical singularity is present there. Now, a conical singularity is a feature that can be tolerated in a number of cases. This is the reason why this condition is analyzed here essentially for completeness in order to be applied wisely to possible relevant configurations, while keeping in mind that any solution which might not satisfy it should anyhow be considered as a proper GR solution.

For cylindrical symmetry and the coordinate frame retained here, this condition can be written as \cite{D06}
\begin{equation}
\frac{\textrm{e}^{-\mu}l'^2}{4l} \stackrel{0}{=} 1, \label{2.33}
\end{equation}
where we insert (\ref{2.2}) and (\ref{2.8}) so as to obtain
\begin{equation}
\frac{c_l}{4r^{a+1}} \stackrel{0}{=} 1, \label{2.34}
\end{equation}
which enforces $a=-1$ and $c_l=4$, that we substitute into (\ref{2.22}) and obtain
\begin{equation}
c^2 = 4(2+\epsilon \sqrt{2}). \label{2.35}
\end{equation}

Now, we insert the above values for parameters $a$, $c_l$, and $c^2$ into (\ref{2.20}) that gives
\begin{equation}
c_P = \frac{1+\epsilon \sqrt{2}}{4 \kappa}. \label{2.36}
\end{equation}

Moreover, implementing $a=-1$ into (\ref{2.22b}), we obtain
\begin{equation}
h = - \frac{1+\epsilon \sqrt{2}}{3 + \epsilon\sqrt{2}}. \label{2.37}
\end{equation}

Therefore, for the physical applications from which conical singularities, and more generally, angular deficits, are excluded, only two independent parameters are left to define each solution of this class: $c_D$ and $c_k$, and the equation of state is fixed by (\ref{2.37}), where the respective signs of the pressure and of the energy density are determined by the choice of $\epsilon$.

Indeed, from (\ref{2.36}) and (\ref{2.37}), we can draw the conclusion that:

- in the case $\epsilon=+1$, $c_P >0$, and thus $P>0$; $h<0$, and thus $\rho<0$. 

- in the case $\epsilon=-1$, $c_P <0$, and thus $P<0$; $h>0$, and thus $\rho<0$.

In both cases, the weak energy condition is not fulfilled. 

We can therefore conclude that subclass B.2 solutions either are free of angular deficit in the vicinity of the axis but exhibit a non usual equation of state and do not satisfy the weak energy condition, or they present an angular deficit but they can provide a wider range of physical possibilities. It is this range which is explored in the following.

Indeed, the constraints analyzed above are more or less mandatory. The metric signature and the axisymmetry conditions must obligatorily be fulfilled by the solutions so that they can be considered as actual GR solutions. Conversely, the junction conditions and the energy conditions discussed above, even if necessary in a standard astrophysical context, are not fundamental to validate the solutions as genuine GR ones.

\subsubsection{Hydrodynamical properties}

Now, we calculate the hydrodynamical properties of the fluid, using the implementation of the well-known formalism to the cylindrically symmetric case as completed  by C\'el\'erier and Santos \cite{CS20}. The considered solutions are those exhibiting sufficient matter inside, i. e., infinitely wide cylinders.

The non-zero component of the acceleration vector can be written as
\begin{equation}
\dot{V}_1= - \Psi, \label{2.38}
\end{equation}
with
\begin{equation}
\Psi= -\frac{1}{2} (v^2 f' - 2 v \Omega k' - \Omega^2 l'), \label{2.39}
\end{equation}
where we insert (\ref{2.2}), (\ref{2.14}), (\ref{2.15}), (\ref{2.18}), and (\ref{2.19}) so as to obtain
\begin{equation}
\Psi= - \left[\frac{1}{2} + \frac{4 c^2}{(a-1)c_l^2} \right] \frac{1}{r}. \label{2.40}
\end{equation}
It comes, therefore,
\begin{equation}
\dot{V}_1= \left[\frac{1}{2} + \frac{4 c^2}{(a-1)c_l^2} \right] \frac{1}{r}. \label{2.41}
\end{equation}
Now, the squared modulus of this acceleration vector reads
\begin{equation}
\dot{V}^{\alpha} \dot{V}_{\alpha}= \textrm{e}^{-\mu} \Psi^2, \label{2.42}
\end{equation}
where we substitute (\ref{2.9}) and (\ref{2.40}), that yields
\begin{equation}
\dot{V}^{\alpha} \dot{V}_{\alpha}= \left[\frac{1}{2} + \frac{4 c^2}{(a-1)c_l^2} \right]^2 \frac{1}{r^{2+a}}. \label{2.43}
\end{equation}

The shear tensor exhibits two non-zero components which verify
\begin{equation}
2 \sigma_{01}= D^2 \Omega(v' \Omega - v \Omega'), \label{2.44}
\end{equation}
\begin{equation}
2 \sigma_{13}= D^2 v (v \Omega' - v' \Omega), \label{2.45}
\end{equation}
where (\ref{2.12}), (\ref{2.18}), and (\ref{2.19}) are inserted so as to give
\begin{equation}
2 \sigma_{01}= - \frac{2c}{\sqrt{c_l r}}\left[c_k + \frac{c_D(1-a)}{2c} r^{ \frac{4c^2}{(a-1)c_l^2}}\right], \label{2.46}
\end{equation}
\begin{equation}
2 \sigma_{13}= - \frac{2c} {\sqrt{c_l r}}. \label{2.47}
\end{equation}
Its squared modulus reads
\begin{equation}
\sigma^2= \frac{\textrm{e}^{-\mu} D^2}{4} (v \Omega' - v' \Omega)^2, \label{2.48}
\end{equation}
which becomes, with (\ref{2.9}), (\ref{2.12}), (\ref{2.18}), and (\ref{2.19}) inserted
\begin{equation}
\sigma^2= \frac{c^2}{c_l^2} \frac{1}{r^{2+a}}. \label{2.48a}
\end{equation}
In accordance with what is expected for a differentially rotating fluid, the shear does not vanish. 

Conversely, since the fluid exhibits an irrotational flux, its rotation tensor and scalar vanish.

\subsubsection{Summary: solutions with an angular deficit}

We display here, for the reader's convenience, a summary of the equations describing the class B.2 solutions. One can easily obtain the solutions with no angular deficit by implementing $a=-1$ and $c_l=4$ and $c^2 = 4 (2 + \epsilon \sqrt{2})$ into the below set of equations that reads
\begin{equation}
f = \frac{c_D^2}{c_l} r^{1 + \frac{8c^2}{(a-1)c_l^2}} - c_l r \left[c_k + \frac{c_D(1-a)}{2c} r^{ \frac{4c^2}{(a-1)c_l^2}}\right]^2, \label{2.49}
\end{equation}
\begin{equation}
\textrm{e}^{\mu}= r^a, \label{2.50}
\end{equation}
\begin{equation}
k = c_l r \left[c_k + \frac{c_D(1-a)}{2c} r^{ \frac{4c^2}{(a-1)c_l^2}}\right], \label{2.51}
\end{equation}
\begin{equation}
l = c_l r, \label{2.52}
\end{equation}
\begin{equation}
h = \frac{1+3a - \epsilon \sqrt{9+10a+9a^2}}{7 + a + \epsilon \sqrt{9+10a+9a^2}}, \label{2.53}
\end{equation}
\begin{equation}
P = \frac{c_P}{r^{\left(\frac{1+h}{h}\right)\left[\frac{1}{2} + \frac{4c^2}{(a-1)c_l^2}\right]}}, \label{2.54}
\end{equation}
\begin{equation}
\rho = \frac{c_P}{h r^{\left(\frac{1+h}{h}\right)\left[\frac{1}{2} + \frac{4c^2}{(a-1)c_l^2}\right]}}, \label{2.55}
\end{equation}
\begin{equation}
\Omega^2 = \frac{c_l \left[c_k + \frac{c_D(1-a)}{2c} r^{ \frac{4c^2}{(a-1)c_l^2}}\right]^2}{c_D^2 r^{1 + \frac{8c^2}{(a-1)c_l^2}}}, \label{2.56}
\end{equation}
\begin{equation}
v = - \frac{\sqrt{c_l}}{c_D}\frac{1}{r^{\frac{1}{2} + \frac{4c^2}{(a-1)c_l^2}}}, \label{2.57}
\end{equation}
\begin{equation}
D = c_D r^{1 + \frac{4c^2}{(a-1)c_l^2}}, \label{2.58}
\end{equation}
with
\begin{equation}
c_P = \frac{1}{16 \kappa} \left[3 + 8a + 9a^2 - \epsilon(1+3a) \sqrt{9+10a+9a^2}\right], \label{2.59}
\end{equation}
\begin{equation}
\frac{c^2}{c_l^2} = \frac{(1-a)}{16} \left(1-3a + \epsilon \sqrt{9+10a+9a^2}\right). \label{2.60}
\end{equation}

\subsubsection{Energy conditions}

It appears from (\ref{2.55}) that the energy density is positive if the $c_P$ and $h$ parameters exhibit the same sign. In this case, the weak energy condition is fulfilled and the corresponding spacetimes are physically well-behaved from a standard fluid point of view.

From (\ref{2.54}), it comes out moreover that the subclass of solutions with $c_P>0$ present a positive pressure, thus satisfying the strong energy condition and improving their physical characteristics. In this case, $h$ must also be positive so as to preserve the fulfilment of the weak energy condition.

\subsection{Subclass B.3} \label{B.3}

This subclass is defined by $\mu'=0$, which implies $\mu = const.$ and therefore
\begin{equation}
\textrm{e}^{\mu} = c_{\mu},\label{3.1}
\end{equation}
where the integration constant $c_{\mu}$ can be set to unity by a rescaling of the $r$ and $z$ coordinates. We have thus
\begin{equation}
\textrm{e}^{\mu} = 1.\label{3.2}
\end{equation}
By inserting $\mu'=0$ and $\mu''=0$ into (\ref{24a}), we obtain
\begin{equation}
\frac{D'}{D} = \frac{l''}{l'}, \label{3.3}
\end{equation}
which can be integrated by
\begin{equation}
D = c_D l'.\label{3.4}
\end{equation}
Then, by completing an analogous insertion into (\ref{25a}), we obtain
\begin{equation}
2ll''-l'^2+4c^2 = 0,\label{3.5}
\end{equation}
whose solution is
\begin{equation}
l= c_l r^2 + 2c r,\label{3.6}
\end{equation}
where $c_l$ is a new integration constant. Substituting $l'$ as obtained from (\ref{3.6}) into (\ref{3.4}) yields
\begin{equation}
D = 2 c_D (c_l r +c).\label{3.7}
\end{equation}

Now, we substitute (\ref{3.6}) and (\ref{3.7}) into (\ref{7}) and obtain, after integration,
\begin{equation}
k = c_k (c_l r^2 +2cr) +2c c_D.\label{3.8}
\end{equation}
Then, $f$ proceeds, from the definition of $D$ given by (\ref{D2}), as
\begin{equation}
f = \frac{4c_D^2(c_l r+ c)^2 - \left[c_k (c_l r^2 +2cr) +2c c_D\right]^2}{c_l r^2 + 2c r}.\label{3.9}
\end{equation}

From (\ref{3.7}), we obtain $D''=0$, which, once inserted into (\ref{12}), gives
\begin{equation}
P= 0.\label{3.10}
\end{equation}

Now, we insert $\mu''=0$, $D$ given by (\ref{3.7}), and derivatives, $f$, given by (\ref{3.9}), and derivatives, $l$, given by (\ref{3.6}), and derivative, into (\ref{G00}) where we have inserted (\ref{7}) and obtain
\begin{equation}
\rho= 0.\label{3.11}
\end{equation}

It appears therefore that subclass B.3 is composed of vacuum spacetimes which  must be discarded in the framework of perfect fluid solutions.

\subsection{Subclass B.4} \label{B.4}

This subclass is defined by $\mu''=0$, which can be integrated as
\begin{equation}
\mu = r +c_{\mu}, \label{4.1}
\end{equation}
after a rescaling of the $r$ and $z$ coordinates.

Now, by inserting $\mu$, as given by (\ref{4.1}), and derivatives into (\ref{12b}), (\ref{14}) and (\ref{25a}) we obtain
\begin{equation}
\frac{D''}{D} = \left(\frac{3}{2} + \frac{l'}{2l}\right) \frac{D'}{D} - \frac{l''}{2l}, \label{4.2}
\end{equation}
\begin{equation}
\frac{D'}{D} = \frac{\left(\frac{l''}{l} - \frac{l'^2}{l^2} + \frac{4c^2}{l^2} \right)}{1 - \frac{l'}{l}}, \label{4.3}
\end{equation}
and
\begin{equation}
\frac{l''}{l}\left[(1-h)\frac{l'}{l} +1+h\right] + \left(\frac{2c^2}{l^2} - \frac{l'^2}{2l^2}\right)\left[(1-h)\frac{l'}{l} + 3+h\right] =0, \label{4.4}
\end{equation}
which constitute a set of three equations for three unknown functions of $r$. However, this set of equation is not solvable analytically, even for an equation of state of the form $h=const.$ For this last case, however, (\ref{4.4}) might be solved numerically, giving, therefore, $l(r)$. By inserting this metric function and its derivatives into (\ref{4.3}), we can obtain $D(r)$, still numerically. Then, $k(r)$ would proceed from (\ref{7}), and $f(r)$, from (\ref{D2}). Finally $P(r)$, and then $\rho(r)=P(r)/h$, would be obtained from (\ref{12}). The global rotation parameter $\Omega$ and the velocity $v$ arise from (\ref{timelike}) and (\ref{irr}).

\subsection{Subclass B.5} \label{B.5}

This subclass is defined by $h=1$, which implies $P=\rho$. This definition inserted into (\ref{24a}) yields
\begin{equation}
\frac{D'}{D} = -\frac{\mu''}{\mu'}, \label{5.1}
\end{equation}
which can be integrated by
\begin{equation}
D = \frac{c_D}{\mu'}, \label{5.2}
\end{equation}
where $c_D$ is an integration constant.

Now, the Bianchi identity (\ref{Bianchi}), with $h=1$ substituted, becomes
\begin{equation}
\frac{P'}{2P} - \frac{l'}{2l} + \frac{D'}{D} = 0, \label{5.3}
\end{equation}
which can be integrated by
\begin{equation}
D = c_B \sqrt{\frac{l}{P}}, \label{5.4}
\end{equation}
where $c_B$ is an integration constant and which, owing to (\ref{5.2}), can be also written as
\begin{equation}
P = \frac{c_B^2}{c_D^2} l \mu'^2. \label{5.5}
\end{equation}

Then, (\ref{5.1}) inserted into (\ref{14}) yields
\begin{equation}
\frac{\mu''}{\mu'} - \frac{l''}{l'} + \frac{l'}{l}  - \frac{4c^2}{ll'}= 0. \label{5.6}
\end{equation}
Now, (\ref{5.6}), multiplied by $l'/l$, is substituted into (\ref{8}), together with (\ref{5.1}), which gives
\begin{equation}
\frac{f'l'+ k'^2}{2 D^2} = - \left(\frac{\mu'' l'}{2 \mu' l} +   \frac{l''}{2l} \right). \label{5.7}
\end{equation}
Finally, (\ref{5.1}), (\ref{5.5}), and (\ref{5.7}) are substituted into (\ref{G11}) so as to obtain
\begin{equation}
-\mu'' \left(1+ \frac{l'}{2 \mu' l} \right) - \frac{l''}{2l} = 2\kappa  \frac{c_B^2}{c_D^2} l \mu'^2 \textrm{e}^\mu. \label{5.8}
\end{equation}

We have thus a couple of equations involving $l$ and $\mu$. But these equations also cannot be integrated analytically. However, numerical integrations could be performed such that, once $l$ and $\mu$ are found, $D$ proceeds from (\ref{5.2}). Then $k$ is obtained through (\ref{7}). Then $f$ follows fom (\ref{D2}), and $P=\rho$ result from (\ref{5.5}). The global rotation parameter $\Omega$ and the velocity $v$ are still obtained from (\ref{timelike}) and (\ref{irr}).

\subsection{Subclass B.6} \label{B.6}

This subclass is defined by $h=-1$.

Owing to its general form displayed as Eq. (57) in C\'el\'erier and Santos \cite{CS20}, the Bianchi identity can be written here as
\begin{equation}
P' -(\rho +P) \Psi = 0,\label{6.1}
\end{equation}
where we insert $\rho=P/h$ and obtain
\begin{equation}
P' -\left(1 +\frac{1}{h}\right) \Psi = 0,\label{6.2}
\end{equation}
where we substitute $h=-1$, which yields
\begin{equation}
P' = 0.\label{6.3}
\end{equation}
The pressure $P$ is therefore constant and this constant vanishes due to the junction condition on the boundary surface implying P=0 there, and thus everywhere.

Now, since the definition of this subclass yields $\rho=-P$, the energy density $\rho$ vanishes too and the spacetime is a vacuum, not a perfect fluid interior.

\section{Comparison with the rigid rotation case} \label{rr}

In a preceding paper \cite{C23a} the interior spacetimes sourced by stationary rigidly rotating cylindrical perfect fluids have been studied. A class of such solutions has been fully integrated, and its properties thoroughly examined. It is therefore interesting to contrast these results with those displayed here as subclass B.2.

\subsection{Number of degrees of freedom}

In both cases, the number of unknown functions to be calculated is larger than the number of field equations to be solved. The difference lies in the number of degrees of freedom left. They have been addressed in two different ways.

In the rigid case, where the considered class of solutions emerges quite naturally from the field equations, no further constraint has been necessary to be able to solve the problem. Since this class is only a partial set of the global solutions, and since we had only one degree of freedom to deal with, the choice of this peculiar class was enough to close the set of equations and their integration has been performed without the need for any other assumption.

In the case of differential rotation, the issue is more tricky, since we are confronted to two extra-degrees of freedom. We have therefore chosen, as a first assumption, to consider an irrotational fluid, since this choice provides a partial integration of some of the equations. As regards the second degree of freedom, it has been naturally dealt with by the form of the equations itself. Indeed, seven different classes and subclasses have emerged from their arrangement. The consideration, in turn, of each of them has therefore closed the set.

\subsection{The integrated classes}

For rigid rotation, only one class of two parameter solutions has been fully integrated and analyzed, exhibiting all the demanded properties of well-behaved spacetimes. Moreover, for some achievable parameter ranges, they can satisfy the weak or/and the strong energy conditions. Their matching to a vacuum exterior has been easily completed. These solutions provide, therefore, a set of interesting spacetimes for astrophysical applications.

The outcome of the study of differentially rotating fluids is both more complete and more complex. We have identified and examined from an analytical point of view seven classes or subclasses to the irrotational fluid problem. Among this bunch of classes, some have been ruled out as either representing a vacuum or being only numerically integrable. However, we have sorted out two classes which have been fully analytically integrated: class A, which appeared, when matched to a vacuum exterior, as a dust solution with negative energy density and singular axis, and subclass B.2, representing spacetimes sourced by a proper perfect fluid exhibiting a polytropic equation of state with $h=const.$

\subsection{Matching to a vacuum exterior}

As we have recalled above, the matching of spacetimes generated by a rigidly rotating source arises naturally and does not imply any bad consequence for the interior cylinder.

Instead, in the differential rotation case, the junction conditions compel the interior spacetimes to exhibit unwanted features. Class A solutions become dust filled, subclasses B.3 and B.6 turn to vacuum and the subclass B.2 cylinders become thread-like or infinitely wide, depending on the respective values of parameters $h$ and $a$ of the solutions, or possibly vacuums for very special values of these parameters. 

As suspected, the matching appears therefore as a difficult operation when the fluid is differentially rotating, while it occurs naturally when the rotation is rigid.

\subsection{Equation of state}

For a rigidly rotating fluid, the equation of state emerges naturally from the calculations. Its departure from that of an ultra-relativistic gas can be adjusted at will by a tuning of the parameters of the solution. However, the nature of the equation of state cannot be any.

Differential rotation, instead, allows a wider choice of equations of state. Leaving out the dust and vacuum cases, we focus our interest on subclass B.2. It has been integrated for the special case where $h=const.$, and allows the choice of a wide set of polytropic equations of state, which are popular for a number of astrophysical applications. This is an advantage, if it were not for the problem with the matching.

\subsection{Angular deficit}

The so-called ``regularity'' condition, which is, as we have shown, a mere condition for avoiding an angular deficit near the axis of rotation, has been imposed without problem to the parameters of the rigidly rotating spacetimes. It yields a constraint affecting the parameters and decreases the number of independent ones.

Now, when applied to the differentially rotating subclass B.2, it yields a negative energy density, and thus, the weak energy condition is no more fulfilled. However, the main consequence of relaxing this condition and, thus, of allowing an angular deficit to be present near the axis, is the occurrence of a conical singularity, which is not too severe a drawback.

\section{Conclusion} \label{concl}

The field equations pertaining to interior spacetimes sourced by stationary differentially rotating cylindrical irrotational perfect fluids have been fully integrated and the mathematical and physical properties of the solutions have been examined.
Directly issuing from the field equations, two main classes have been distinguished. The first, denoted class A, appears, when matched to a vacuum exterior, as generated by a dust source. This source exhibits a negative energy density and a singular axis which could constitute physical drawbacks for most of astrophysical applications. Now, it must be noted that another complete solution for a differentially rotating cylindrical dust source had been proposed long ago by Maitra for the case where the fluid is rotational, i. e., exhibits non vanishing rotation with respect to the velocity vector of the matter \cite{M66}, contrary to the present assumption of irrotational fluid. In Maitra's solution, the matter is distributed with positive energy density \cite{B09}, but the junction conditions are not imposed. The solution is ``open in all spatial direction, i.e. it extends to infinite proper distance in all directions''. This suggests that differential rotation can be exhibited by a stationary dust cylinder of standard matter, i. e., with the weak energy condition fulfilled, provided that the fluid, besides being rotational -- i. e., its ``local'' rotation with respect to its velocity four-vector is nonzero -- is not matched to a vacuum exterior.

The second class of mathematically coherent solutions, denoted class B, has naturally divided into six subclasses. Subclass B.1 must be ruled out since its $l$ metric function vanishes for all $r$. Subclasses B.3 and B.6 have appeared to be a vacuum, not a perfect fluid solution, and have thus been dismissed. Subclass B.4 and Subclass B.5 have been partially integrated and, even if a full analytical solution has not been found, the equations involving the metric functions have been proposed as a seed for future numerical integrations. Summarized integration methods have been displayed for such a purpose.

Subclass B.2 is the most interesting. It is composed of analytical solutions where the ratio $P/\rho=h$ is a constant, i. e., the equation of state is polytropic. Its different mathematical and physical properties have been examined. The hydrodynamic properties of the fluid have been calculated. This has allowed to verify that the shear is nonzero while the rotation vanishes, which was indeed expected from a differentially rotating source. We have shown that a proper signature of the metric can occur for a large range of values for the four independent parameters of the solutions. The junction conditions with an exterior Lewis-Weyl vacuum have been analyzed. The results, summarized in Tables I and II, imply that a properly matched spacetime of this subclass is either an infinite thread of zero width whose interior possesses no physical meaning, or an infinitely wide cylinder, or a vacuum. The matching procedure is therefore very tricky here. The condition for avoiding any angular deficit implies also unwanted features for the fluid since the weak energy condition cannot be fulfilled if this condition is satisfied. This four-parameter class of exact analytical solutions must therefore be considered as a set of genuine GR solutions whose use for physical applications is restricted to some particular configurations. For instance, an infinitely long thread-like spacetime can be used to approximate cosmic topological defects, such as strings or superstrings, or, possibly, straight parts of the cosmic web. An infinitely wide cylinder might be considered as cosmological, but its interpretation is unclear.

However, we want to recall here Griffiths and Poldosk\'y's remark \cite{G09}: ``much can be learned about the character of gravitation and its effects by investigating particular idealised examples.''

\end{document}